\documentclass[fleqn,usenatbib]{mnras}

\newcommand{\kjmol}{kJ mol$^{-1}$}

\usepackage{graphicx} 
\usepackage{newtxtext,newtxmath}
\usepackage[english]{babel} 
\usepackage[utf8]{inputenc} 
\usepackage[T1]{fontenc}  
\usepackage{url} 
\usepackage[normalem]{ulem}
\usepackage[version=3]{mhchem}
\usepackage{epstopdf} 
\usepackage{multirow}
\usepackage{chemfig}
\usepackage{makecell}
\usepackage{xcolor, soul}
\usepackage{setspace}
\usepackage{booktabs}

\graphicspath{{./}{figures/}}

\title[Hydrogenation of CS$_2$]{Experimental and computational studies of the hydrogenation of carbon disulfide (\ce{CS2}) on ice analogues}

\author[T. Nguyen et al.]{
Thanh Nguyen$^{1}$\thanks{E-mail: tnguyen@strw.leidenuniv.nl} \thanks{Current address: Laboratory for Astrophysics, Leiden Observatory, Leiden University, PO. Box 9513, NL 2300 RA Leiden, The Netherlands.},
Germ\'an Molpeceres$^{2}$,
Yasuhiro Oba$^{1}$,
Marcelino Ag\'undez$^{2}$, 
Gisela Esplugues$^{3,4}$,
\newauthor Jos\'e Cernicharo$^{2}$,
and Naoki Watanabe$^{1}$
\\
$^{1}$ Institute of Low Temperature Science, Hokkaido University, Kita 19, Nishi 8, 060-0819, Sapporo, Japan\\
$^{2}$Instituto de Fisica Fundamental, CSIC, Calle Serrano 123, E-28006, Madrid, Spain \\
$^{3}$ Observatorio Astronómico Nacional (OAN), Alfonso XII, 3, 28014 Madrid, Spain \\
$^{4}$ Observatorio de Yebes, IGN, Cerro de la Palera s/n, E-19141 Yebes, Guadalajara, Spain
}

\date{Accepted 2026 February 23. Received 2026 February 23; in original form 2025 December 21}

\pubyear{\the\year{}}
\begin{document}
\label{firstpage}
\pagerange{\pageref{firstpage}--\pageref{lastpage}}
\maketitle

\begin{abstract}
Carbon disulfide (CS$_2$) is one of the sulfur-bearing species expected to be present in the interstellar medium (ISM). In this study, we investigated the surface reactions of solid CS$_2$ with hydrogen (H) atoms on amorphous solid water (ASW) using laboratory experiments supported by computational calculations. Our results show that CS$_2$ reacts with H atoms through quantum tunneling in the initial step, followed by successive H addition reactions, with or without activation barriers, on icy surfaces. These processes lead to the formation of several sulfur-bearing species, including hydrogen sulfide (H$_2$S), methyl mercaptan (CH$_3$SH), and small amounts of dithioformic acid (HC(S)SH) and methanedithiol (CH$_2$(SH)$_2$). The observed reactivity of CS$_2$ with H atoms provides a plausible explanation for the non-detection of CS$_2$ in interstellar ices. Furthermore, the efficient hydrogenation of the complex molecules derived from CS$_2$, namely HC(S)SH and CH$_2$(SH)$_2$, suggests that these species could be easily undergone with H atoms to produce other S-bearing species under ISM conditions. 
\end{abstract}




\section{Introduction}

 In recent years, sulfur (S) chemistry on icy grains mantles in dense clouds, prestellar cores, cometary ices, and planets has attracted increasing attention not only from the astrochemical community but also from the fields of astrobiology and planetary science \citep{Mifsud2021}. To date, more than 30 S-bearing species have been identified in the interstellar medium (ISM), and most of which have been detected only in the gas phase. Recent observations from the \textsc{Quijote} line survey have identified many S-bearing species in dense clouds, such as NCS, HCCS, H$_{2}$CCS, H$_{2}$CCCS, C$_{4}$S, C$_{5}$S, HCSCN, HCSCCH, and NCCHCS \citep{Cernicharo2021d,Cernicharo2021e, Cabezas2024}, NC$_{3}$S, HC$_{3}$S, and HNCS \citep{Cernicharo2024b,Cernicharo2024a}, CH$_{3}$CHS and CH$_{2}$CHCHS \citep{Agundez2025, Cabezas2025}. However, the total observed abundance of S-bearing species in dense clouds remains significantly lower than the expected cosmic abundance of sulfur \citep{Wakelam2004a, Anderson2013, Fuente2023}. This mismatch is commonly referred to as the “sulfur depletion problem” \citep{Ruffle1999, Vidal2017, Laas2019}, in which the “missing sulfur” is thought to be trapped within interstellar grains. The mechanisms driving the chemical conversion of sulfur-bearing species on grain surfaces remain poorly understood, hindering the identification of the specific compounds likely to be present in interstellar ices. Laboratory experiments and computations have investigated surface processes for S-bearing species in the ISM. Some of these studies include chemical pathways starting from hydrogen sulfide (H$_2$S) and methyl mercaptan (CH$_3$SH) \citep{Lambert2017, Oba2018, Lamberts2018, Nguyen2023}. These two species are considered to form on dust grains \citep{Millar1986, Millar1990, Lamberts2018}, they have been detected only in the gas phase, suggesting that they are desorbed into the gas phase or transformation into other S-bearing species by (non-)energetic processes in interstellar dust grains, as indicated in previous studies \citep{Ferrante2008,Garozzo2010,Jimenez-Escobar2014, Oba2018, Nguyen2023}. In addition to H$_2$S and CH$_3$SH, carbonyl sulfide (OCS) and sulfur dioxide (SO$_2$) have been identified as dominant sulfur-bearing species in the solid phase. Several laboratory studies have demonstrated their formation via ion irradiation of S-containing species mixed with ice components or through S-ion implantation into O-bearing species (e.g., O$_2$, CO, and CO$_2$) \citep{Ferrante2008, MIFSUD2024}. However, their abundances are estimated to be significantly lower than the total expected sulfur abundance, implying that the missing sulfur problem is present in icy grain mantles. A suggestion is that a substantial fraction of sulfur may be locked in refractory reservoirs or converted into other S-bearing species through (non-)energetic processing \citep{Shingledecker2020, Mifsud2021, Mifsud2022_2}. Indeed, surface reactions of OCS and SO$_2$ with H atoms have been investigated under conditions in the ISM, suggesting the formation of some S-bearing species such as HCSOH, HSOOH, and HOSOH, which, however, have not been detected yet in the ISM \citep{Molpeceres2021,Molpeceres2022_thio, Nguyen2021, Nguyen2024}. Some characteristic features of S-bearing molecules were found in previous laboratory studies \citep{Nguyen2021, Nguyen2023, Nguyen2024}, where S-bearing species including OCS, \ce{SO2}, and \ce{CH3SH} yielded \ce{H2S} as a final product after reacting with H atoms on interstellar ice analogs at low temperatures. After formation, \ce{H2S} can be easily desorbed into the gas phase by chemical desorption with the efficiency of 3 $\pm$ 1.5$\%$ per reaction \citep{Oba2018, Furuya2022}. Additionally, the absorption coefficient of H$_2$S  has been investigated to be low in dense clouds and solid H$_2$S can further react with other species on dust grains (e.g., NH$_3$) to form more complex molecules (e.g., ammonium salts, NH$_4$SH; \cite{Slavicinska2025}). Alternatively, H$_2$S can be converted to the high-molecular-weight S-containing molecules, including sulfanes (H$_2$S$_n$; n = 2-11) and octasulfur (S$_8$), by galactic cosmic ray processing on icy dust grains \citep{Herath2025}. These processes can collectively contribute to explaining the "sulfur depletion problem" in dense clouds. The preferential formation of \ce{H2S}, a simpler molecule than its precursors \citep[e.g., OCS, CH$_3$SH, SO$_2$;][]{Nguyen2021, Nguyen2023, Nguyen2024}, together with its potential for chemical desorption, is intriguing and may partly account for the missing sulfur in dark clouds.

In this context, carbon disulfide (CS$_2$) emerges as a promising candidate for advancing our understanding of the resilience of sulfur-bearing molecules on interstellar ices. The existence of CS$_2$ in cometary ices has been proposed based on the detection of a high abundance of the CS molecule in comets, which is related with CS$_2$ \citep{Despois1992, Feldman1980, Smith1980, WEAVER1981}. For instance, the abundance of CS$_2$ has been estimated to be approximately 0.2$\%$ relative to H$_2$O in comet C/1995 O1 (Hale-Bopp) \citep{Bockelee2000}. With the \textit{James Webb Space Telescope} (JWST), CS$_2$ has been detected in the atmosphere of the warm sub-Neptune TOI-270 d, in conjunction with other sulfur-bearing species, including CS and H$_2$CS \citep{Moses2024, Felix2025}. The detection of \ce{CS2} on grains in low-temperature regions may provide valuable insights into S chemistry and could also be significant in less evolved regions, such as dark clouds. However, to date, neither \ce{CS2}, which lacks a permanent dipole moment, nor its protonated form, \ce{HSCS+}, have been identified in the interstellar medium (ISM). Alternately, although the formation of CS$_2$ is proposed less effective in dense quiescent cloud, several previous studies have suggested that the lack of observational detection of CS$_2$ may result from the overlap of its infrared absorption bands with those of other species and/or its rapid conversion into other products shortly after formation \citep{Garozzo2010}. Therefore, laboratory investigations of the chemical reactivity of CS$_2$ on interstellar dust grains are essential for understanding its astrochemical role.

Laboratory experiments indicated that CS$_2$ can be formed by ion irradiation of mixed ices containing H$_2$S with CO or by He$^{+}$ ion irradiation of ice components (e.g., CO, CO$_2$) deposited on top layers of allotropic sulfur \citep{Garozzo2010, Mifsud2025}. Additionally, CS$_2$ formation has been reported following proton irradiation of OCS \citep{Ferrante2008}, as shown in the reactions below ocurring on ices: 
\begin{align}
    \ce{OCS -> CO + S},\\
    \ce{S + OCS -> OCS2},\\
    \ce{OCS2 -> O + CS2}.
    \label{rct:CS2_f}
\end{align}
Also, astrochemical models postulate that the main contributing grain phase formation routes are \citep{Laas2019}:
\begin{align}
    \ce{C + S2 &-> CS2}, \\
    \ce{S + CS &-> CS2}, \\
    \ce{S + HCS &-> CS2 + H}.
\end{align}

Among these reactions, it is worth noting that those involving atomic carbon are likely less efficient, since atomic carbon is predominantly chemisorbed on dust grains \citep[occupying more than 70$\%$ of binding sites according to][]{Molpeceres2021c, tsuge_surface_2023}. However, in contrast to its formation processes, the reactions of \ce{CS2} with other species have not been examined in detail, with, to the best of our knowledge, only a few studies addressing this topic through electron bombardment and thermal reaction experiments \citep{Bohn1992, Tsuge2017, Ward2012}. More recently, the chemical behavior of CS$_2$ under energetic irradiation has been explored in laboratory experiments \citep{Rafael2024, Basalgete2026}.

In this study, we focus on the chemical reactions of solid CS$_2$ with H atoms on an icy surface at low temperatures (typically 10~K) using experimental methods supported by preliminary theoretical calculations to understand the physicochemical behavior of \ce{CS2} in interstellar ice. 

\section{Experimental procedure}

All experiments were performed using the Apparatus for Surface Reaction in Astrophysics (ASURA). The ASURA system described in detail \citep{WATANABE2006, Nagaoka2007, Nguyen2020, Nguyen2021, Nguyen2023, Nguyen2024}. In brief, it is composed of an ultrahigh vacuum chamber with a base pressure of 1 $\times$ 10$^{-8}$ Pa, an aluminum (Al) substrate mounted on a He cryostat, an atomic source, a Fourier-transform infrared spectrometer (FTIR), and a quadrupole mass spectrometer (QMS). The surface temperature can be controlled from 5 to 300~K. 

The chemical reactions of CS$_2$ with hydrogen atoms were investigated on a compact amorphous solid water (c-ASW) substrate at low temperatures (typically 10 K). The c-ASW was prepared by depositing H$_2$O vapor onto the surface at 110 K. The thickness of the c-ASW layer was estimated to be 20 monolayers (ML;~1~ML~=~1~$\times$~10$^{15}$~molecules~cm$^{-2}$). After that, the surface temperature was lowered to 10 K for the investigation of the hydrogenation of CS$_2$. The number of monolayers corresponding to the H$_2$O ice thickness is estimated using the following equation (\ref{eq:abs}):

\begin{equation}
    \rm{N = 2.3 \times \frac{sin(7) \times \int Abs(\nu)d\nu}{R \times A} }
    \label{eq:abs}
\end{equation}
where $\int$Abs($\nu$)d$\nu$ is the integrated absorbance of O-H stretching band at 3300 cm$^{-1}$, and A is the apparent band strength for transmission infrared spectroscopy (in molecule$^{-1}$cm), taken from the literature \citep[A = 2 $\times$ 10$^{-16}$ molecule$^{-1}$cm,][]{Gerakines1995}. For our FTIR setup, the correction factor, R, accounting for the conversion from transmission to reflection spectroscopy, was estimated to be 2.

Gaseous CS$_2$ was pre-deposited onto the c-ASW substrate at 10~K with a deposition rate of 1 ML minute$^{-1}$, resulting in an abundance of approximately 1 ML of CS$_2$. This was determined using the absorbance area of the C-S stretching band at 1540 cm$^{-1}$ and a band strength of 9.13 × 10$^{-17}$ molecule$^{-1}$ cm \citep{Garozzo2008}. The pre-deposited CS$_2$ was exposed to hydrogen (H) or deuterium (D) atoms for up to 2 hours to investigate the reaction kinetics. In a separate experiment, the gaseous CS$_2$  was co-deposited with H atoms on the c-ASW substrate at 10 K to increase the yields for better identifying the potential products from the chemical reactions of CS$_2$ with H atoms, and the deposition rates of CS$_2$ were approximately 0.08 or 1~ML min$^{-1}$ during the co-deposition. 

H (or D) atoms were generated by dissociating H$_2$ (D$_2$) in a microwave-discharged plasma in the atomic source. The H (D) atoms were then cooled to 100~K through multiple collisions with the inner wall of an Al pipe connected to the atomic source, which was maintained at 100~K \citep{Nagaoka2007}. The fluxes of H and D atoms were estimated to be 1.4 × 10$^{14}$ and 2.3 × 10$^{14}$ cm$^{-2}$ s$^{-1}$, respectively, using the method outlined by \cite{Oba2014}. 

The chemical reactions of CS$_2$ with H (D) atoms were monitored in situ using FTIR with a resolution of 2~cm$^{-2}$. The reactants and products desorbed from the substrate were detected with the QMS through temperature programmed desorption (TPD) method at the ramping rate 4~K~minutes$^{-1}$.

\section{Results and discussions}
\subsection{Preliminary computational exploration of the \ce{CS2} hydrogenation mechanism} \label{sec:comp}
\begin{figure*}
\centering
	\includegraphics[width=\textwidth]{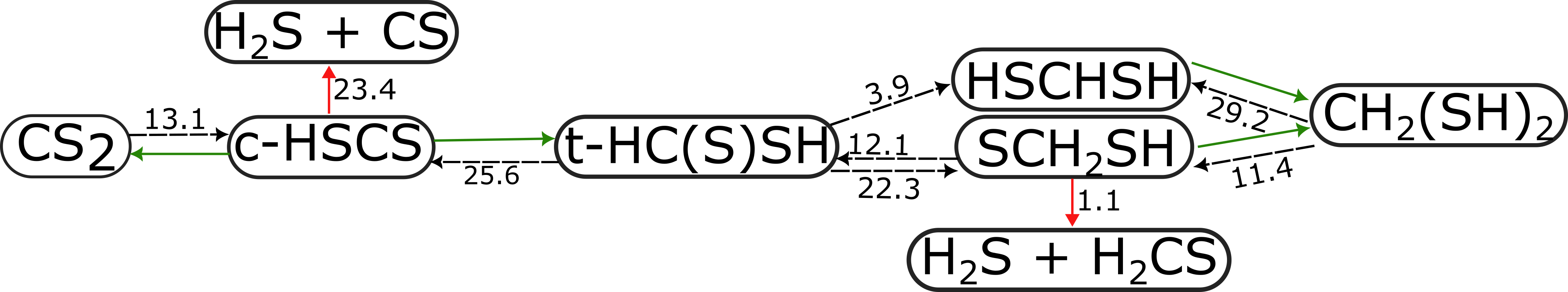}
\caption{Summary of the simplified reaction network obtained in this work. Red arrows indicate irreversible pathways, green arrows indicate barrierless pathways and black dashed lines indicate reactions with a barrier, with the number representing the theoretically derived activation energy. Reactions from right to left occur with release of \ce{H2} (hydrogen abstractions). Energies are shown in \kjmol.}
\label{fig:summary}
\end{figure*}

\begin{table}
\begin{center}
	\caption{Reaction energies ($\Delta U_{\rm R}$) and activation energies ($\Delta U_{\rm A}$) of the simplified reaction network presented in Figure \ref{fig:summary}. All energies are presented in \kjmol. BL means barrierless. In barrierless reactions, the reaction energy is calculated from the energies of the separated reactants as $\Delta U_{R}$ = $U_{AB}$ - ($U_{\rm A}$ + $U_{\rm B}$) and in reactions with a barrier $\Delta U_{\rm R}$ = $U_{\rm AB}$ - ($U_{\rm PRC}$) and $\Delta U_{\rm A}$ = $U_{\rm}$ - ($U_{\rm PRC}$), with PRC meaning pre-reactant complex and TS transition state.  }
	\label{tab:summary}
	\begin{tabular}{c|cc}
	\toprule
	Reaction & $\Delta U_{R}$ & $\Delta U_{A}$  \\
	\midrule
    \multicolumn{3}{c}{\textit{First hydrogenation}} \\
	\midrule
    \ce{CS2 + H -> c-HSCS} & -68.5 & 13.1 \\
    \midrule
    \multicolumn{3}{c}{\textit{Second hydrogenation}} \\
    \midrule

    \ce{c-HSCS + H -> t-HC(S)SH} & -364.1 & BL \\
    \ce{c-HSCS + H -> CS2 + H2} & -357.8 & BL \\
    \ce{c-HSCS + H -> H2S + CS} & -222.4 & 23.4 \\

    
    \midrule
    \multicolumn{3}{c}{\textit{Third hydrogenation}} \\
    \midrule

    \ce{t-HC(S)SH + H -> HSCHSH} & -156.4 &   3.9 \\
    \ce{t-HC(S)SH + H -> HSCH2S} & -164.0 & 22.3 \\
    \ce{t-HC(S)SH + H -> c-HSCS + H2} & -61.2 & 25.6 \\



    \midrule
    \multicolumn{3}{c}{\textit{Fourth hydrogenation}} \\
    \midrule
    \ce{HSCHSH + H -> CH2(SH)2} & -366.6 & BL \\
    \ce{SCH2SH + H -> CH2(SH)2} & -353.6 & BL  \\
    \ce{SCH2SH + H -> t-HC(S)SH + H2} & -258.1 & 12.1 \\
    \ce{SCH2SH + H -> H2S + H2CS} & -280.3 & 1.1 \\
    
    \midrule
    \multicolumn{3}{c}{\textit{Fifth hydrogenation}} \\
    \midrule

    \ce{CH2(SH)2 + H -> HSCHSH + H2} & -57.7 & 29.2 \\
    \ce{CH2(SH)2 + H -> SCH2SH + H2} & -64.6 & 11.4 \\
    
    \bottomrule
	\end{tabular}
	\end{center}
	\end{table}

Before examining the experimental evidence for \ce{CS2} hydrogenation, we first performed a preliminary assessment of the most likely reaction pathways for this system. To this end, we carried out quantum chemical calculations using automated reaction discovery and simplified gas-phase models. The gas-phase approach serves as a first-order approximation to reactions occurring on ice surfaces, where reaction barriers can be modulated, as demonstrated in \citet{molpeceres_carbon_2024}. In that work, we showed that for a reaction network comprising 39 reactions, the deviation in activation barriers introduced by representing surface processes with a gas-phase model was, in the vast majority of cases, below 2 kcal mol$^{-1}$. This small discrepancy arises because, in such reactions, the ice matrix primarily acts as a molecular scaffold rather than as a true catalyst. Although the gas-phase model remains an approximation, its use is justified here by our goal of interpreting the experiments and by the inherent complexity of the reaction network. A more quantitative treatment is left for future dedicated work. We specifically employ the multicomponent artificial induced reaction method (MC-AFIR) using the \textsc{Grrm23} \citep{maeda_afir_2013, maeda_toward_2023, grrm23} coupled with the \textsc{Orca6.0.0} code \citep{neese_software_2022} using the $\omega$B97M-V functional \citep{mardirossian__2016} using the def2-TZVPD basis set \citep{rappoport_property-optimized_2010}. All calculations employ a $\gamma$ collision parameter (see \citealt{maeda_afir_2013}) of 250 kJ mol$^{-1}$ that roughly corresponds to the maximum energy barrier that could be overcome in the calculation. We studied the possible formation pathways sequentially, that is, we first investigated the products of the \ce{CS2 + H} reaction and later the hydrogenation of the products, up to the hydrogenation of \ce{CH2(SH)2}. After the MC-AFIR calculations, the approximated transition states are refined at the same level of theory, following by an intrinsic reaction coordinate (IRC) and optimization of their endpoints to collect the reactants and product states of each reaction to determine the reaction and activation energies $\Delta U_{R}$ and $\Delta U_{\rm A}$.

The reaction network is extremely complex, spanning several tens of stationary points, including isomers, carbenes, conformers or hypervalent compounds. Therefore, for the sake of simplicity, in this work we summarize the results to the direct reaction pathway:

\begin{center}
\ce{CS2 <=>[\mathrm{H}][-\mathrm{H_2}] c-HSCS
    <=>[\mathrm{H}][-\mathrm{H_2}] t-HC(S)SH
    <=>[\mathrm{H}][-\mathrm{H_2}] HSCHSH/SCH2CH
    <=>[\mathrm{H}][-\mathrm{H_2}] CH2(SH)2}.%
\end{center}
We will also highlight specific deviations from this scheme, such as the formation of \ce{H2S + CS} or \ce{H2CS + H2S}, when relevant. A visual representation of the reaction network used to interpret our experimental results is provided in Figure \ref{fig:summary} and Table \ref{tab:summary}. This reaction network successfully reproduces the experimental findings. In the first place, the reaction \ce{CS2 + H -> c-HSCS}, overcoming a small barrier of 13.1 \kjmol, presumably by the action of quantum tunneling  as evinced by the experiments using deuterium shown in Section (\ref{sec:HvsD}). Interestingly, as we observed in \citet{Molpeceres2021b} in the case of the hydrogenation of OCS, the addition product of the reaction yields exclusively the \textit{cis} conformer of the HSCS radical. The formation of S(CH)S, \ce{CS2 + H -> S(CH)S}, is much less effective than that of HSCS, owing to the high barrier ($\sim$41.8~kJ.mol$^{-1}$).

The c-HSCS radical can evolve in three different ways, first it can proceed without a barrier to \ce{CS2} (H-abstraction) or to t-HC(S)SH (H-addition). Additionally, there is a channel with a moderate barrier to \ce{H2S + CS}, that is, an irreversible bimolecular channel. As we will show later, since there are available other low barrier fragmentation reactions we consider that the \ce{c-HSCS + H -> H2S + CS} is not favored against the H-abstraction and H-addition reactions. Like in our previous investigations \citep{Molpeceres2021b, molpeceres_hydrogenation_2025} we confirm that the H-addition of c-HSCS produce exclusively t-HC(S)SH that can continue reaction to two different radicals HSCHSH and \ce{SCH2SH}. While both channels have barriers, the H-addition to the sulfur atom (HSCHSH) is favored over the H-addition to the carbon atom (\ce{SCH2SH}), $\Delta U_{\rm A}$= 3.9 \kjmol against $\Delta U_{\rm A}$= 22.3 \kjmol. It is important to note that 22.3 \kjmol is not extremely high and, as we explain in \citet{molpeceres_hydrogenation_2025}, the diffusion-reaction competition needs to be taken into account to discard possible reaction pathways. However, even if \ce{SCH2SH} is formed in appreciable amounts by this pathway, most of its abundance will come from a posterior reaction (see below). Additionally, we found a significant barrier for the H-abstraction reaction \ce{t-HC(S)SH + H -> c-HSCS + H2} $\Delta U_{\rm A}$= 25.6 \kjmol.

Considering the evolution from HSCHSH our MC-AFIR and subsequent geometry optimizations could only find the formation of methanedithiol (\ce{CH2(SH)2}), proceeding without a barrier. Thus, we can define a \textit{forward} reaction network as:
\begin{align}
    \ce{CS2 + H &-> c-HSCS}, \label{eq:h-cs2}\\
    \ce{c-HSCS + H &-> t-HC(S)SH}, \label{eq:h-hscs}\\
    \ce{t-HC(S)SH + H &-> HSCHSH}, \label{eq:h-hcssh}\\
    \ce{HSCHSH + H &-> CH2(SH)2}. \label{eq:ch2sh2}
\end{align}
that proceeds with low or no barriers. We note that some of these reactions are reiterative to the ones that will be presented in the experimental results section, but here are reproduced accounting for isomerism. 

After formation, H-abstraction of \ce{CH2(SH)2} may proceed via two possible pathways on the substrate. First, H-abstraction can take place at the \ce{-CH2-} moiety and at the \ce{-SH} one. The H-abstraction at the \ce{SH} moiety is clearly favored with a barrier of 11.4 \kjmol against the 29.2 \kjmol needed to abstract in the \ce{-CH2-} group. Therefore, in contrast to what happens in the \textit{forward} reaction network, the \textit{reverse} reaction network initiated by \ce{CH2(SH)2} favors the formation of \ce{SCH2SH}. This radical, contrary to the case of HSCHSH can experience H-abstraction reaction back to t-HC(S)SH, whose main hydrogenation pathways were explained above. More importantly, however, is the fact that \ce{SCH2SH} easily ($\Delta U_{\rm A}$=1.1 \kjmol) breaks irreversibly to form \ce{H2S + H2CS}. Later \ce{H2CS}, through sequential hydrogenation, can easily form \ce{CH3SH} \citep{Lamberts2018}. Therefore, the \textit{reverse} reaction network would predominantly be described by:
\begin{align}
    \ce{CH2(SH)2 + H &-> SCH2SH + H2}, \label{eq:ch2sh2_h} \\
    \ce{SCH2SH + H &-> t-HC(S)SH + H2}, \\
    \ce{SCH2SH + H &-> H2S + H2CS}. \label{sch2sh_h}
\end{align}

The quantum chemical calculations presented here, despite simplified in relation to the total reaction network or choice of structural model, are able to explain the formation of all the closed shell molecules found in the experiment, \ce{H2S}, \ce{H2CS}, \ce{CH3SH}, \ce{HC(S)SH}, and \ce{CHS(SH)2} as we exemplify in the rest of the work.

\subsection{Experimental results}
\subsubsection{Reactions of pre-deposition of CS$_2$ with H atoms on c-ASW at low temperatures} \label{sec:exp}

\begin{figure}
    \centering
    \includegraphics[width=\linewidth]{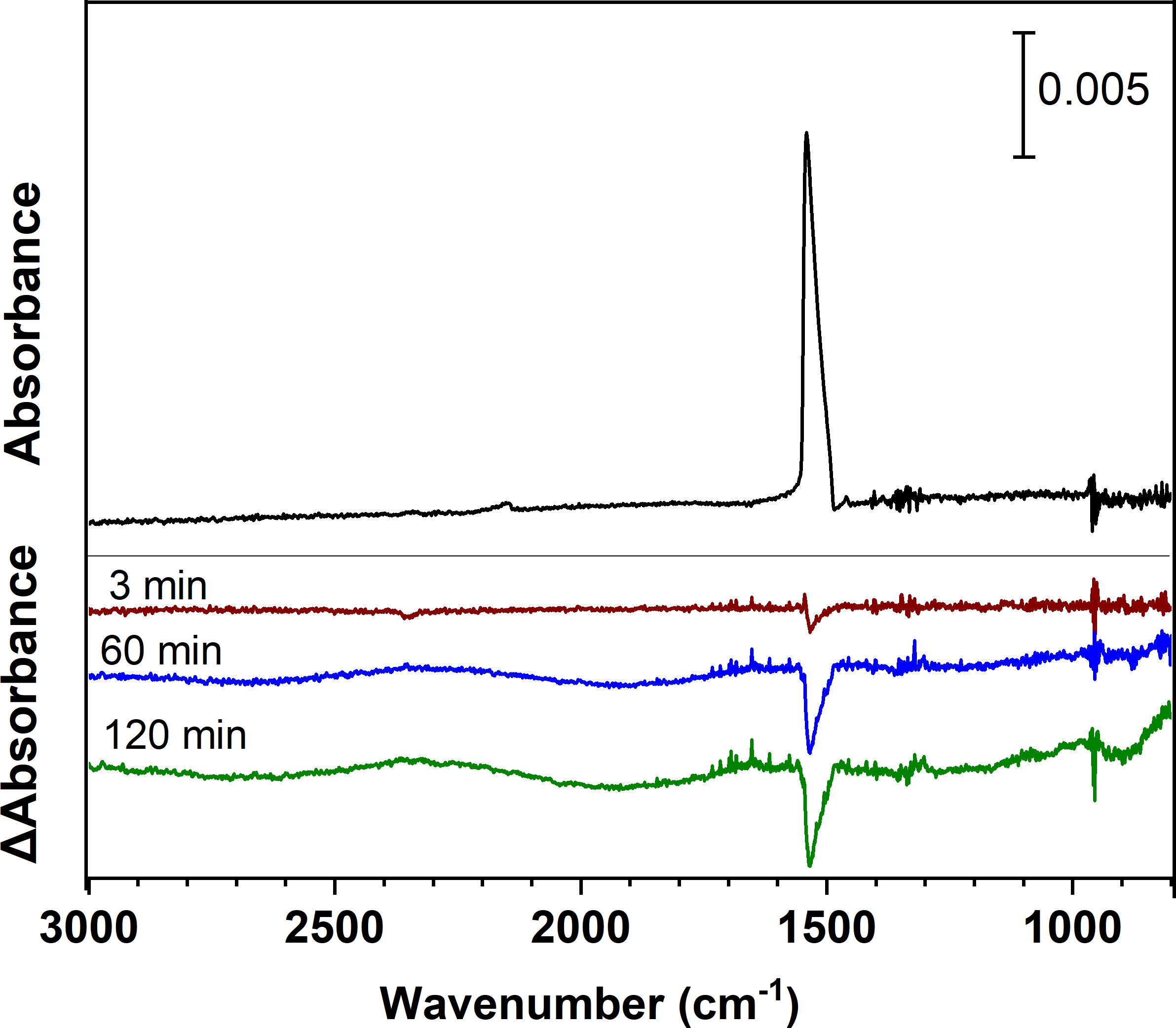}
    \caption{Variation in the different spectra of the solid CS$_2$ after exposure to H atoms for 3, 60, and 120 min at 10~K. A tiny absorption peak observed at 1304 cm$^{-1}$ is assigned to CH$_4$, which may be one of the products formed from reactions of CS$_2$ with H atoms.}
    \label{FTIR_predepositedCS2vsH}
\end{figure}

Figure \ref{FTIR_predepositedCS2vsH} shows the variation in the FTIR spectra of solid CS$_2$ after exposure to H atoms for up to 2 hours, compared to the initial spectrum of unexposed CS$_2$. The observed absorption peak at 1540~cm$^{-1}$ (above the baseline) was attributed to the C-S stretching band of CS$_2$, which aligned well with previous studies \citep{Garozzo2008, Edridge2010}. The decrease in CS$_2$ intensity with exposure to H atoms indicated the loss of CS$_2$ due to reactions with H atoms on the substrate. In contrast, no distinct infrared features of S-containing products were observed, most likely due to the limited number densities of the products.

As discussed in Section \ref{sec:comp}, computational results indicate that reaction (\ref{eq:h-cs2}), in which an H atom adds to the S atom of CS$_2$, has an activation barrier of 13.1 kJ mol$^{-1}$ ($\sim$1575~K). In contrast, the activation barrier for reaction corresponding to H addition at the C atom, is considerably higher, about 41.8~kJ mol$^{-1}$ ($\sim$5027~K). This difference suggests that solid CS$_2$ reacts preferentially with H atoms at the sulfur site, forming the c-HSCS radical. We assume that reaction (\ref{eq:h-cs2}) proceeds at low temperatures with the aid of quantum tunneling, an assumption later confirmed in Section \ref{sec:HvsD}. Once formed, c-HSCS radicals can undergo further hydrogenation to yield new products. In comparison, the destruction to H$_2$S + CS involves moderate activation barriers (Figure \ref{fig:summary}). Moreover, the hydrogen abstraction merely returns to the starting point in the network (\ce{CS2}). These channels should therefore be less favorable than H addition. Consequently, HC(S)SH, produced via reaction (\ref{eq:h-hscs}), is expected to be the predominant product in the continuous hydrogenation of c-HSCS radicals.

Dithioformic acid (HC(S)SH) formed in reaction (\ref{eq:h-hscs}) is expected to further react with H atoms up to the formation of \ce{CH2(SH)2}, which easily fragments to \ce{H2S + H2CS}, and breaking the original \ce{CS2} skeleton. The hydrogenation of H$_2$CS on the surface leads to the formation of CH$_3$SH, consistent with the quantum chemical results reported by \cite{Lamberts2018}, as follows:
\begin{equation}
    \rm{H_2CS} \xrightarrow[]{+H} \rm{H_3CS}/\rm{H_2CSH} \xrightarrow[]{+H} \rm{CH_3SH}.
    \label{rct:4_CH3SH}
\end{equation}

On the other hand, H$_2$S is expected to desorb in the gas phase through chemical desorption \citep{Oba2018}:
\begin{align}
     \ce{H2S + H &-> HS + H2}, \label{rct:5_H2S-HS}\\
     \ce{HS + H &-> H2S(g)}. \label{rct:5_HS-H2S}
\end{align}
where (g) is used to identify the release to the gas phase. 

\begin{figure}
    \centering
    \includegraphics[width=1.1\linewidth]{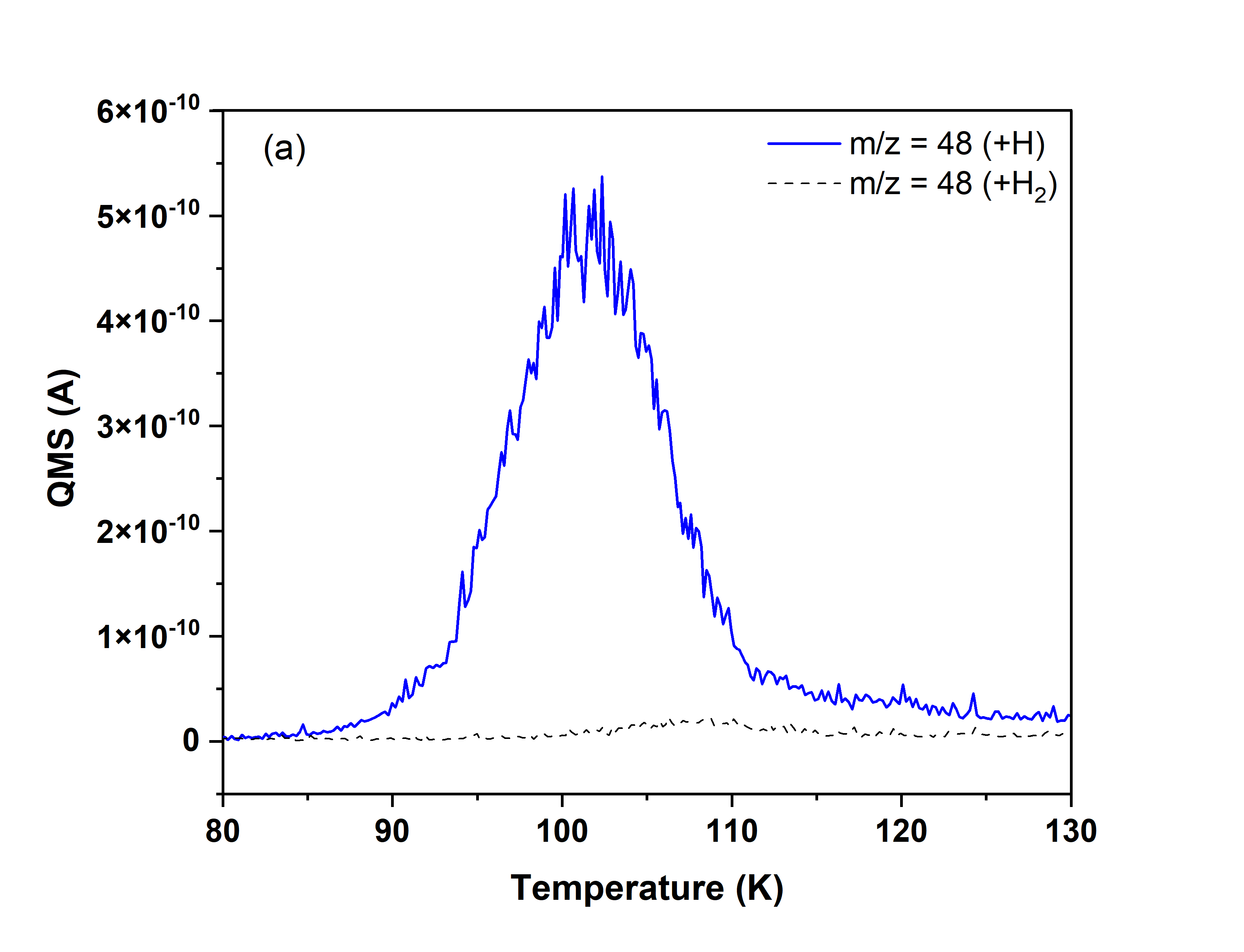}
    \includegraphics[width=1.1\linewidth]{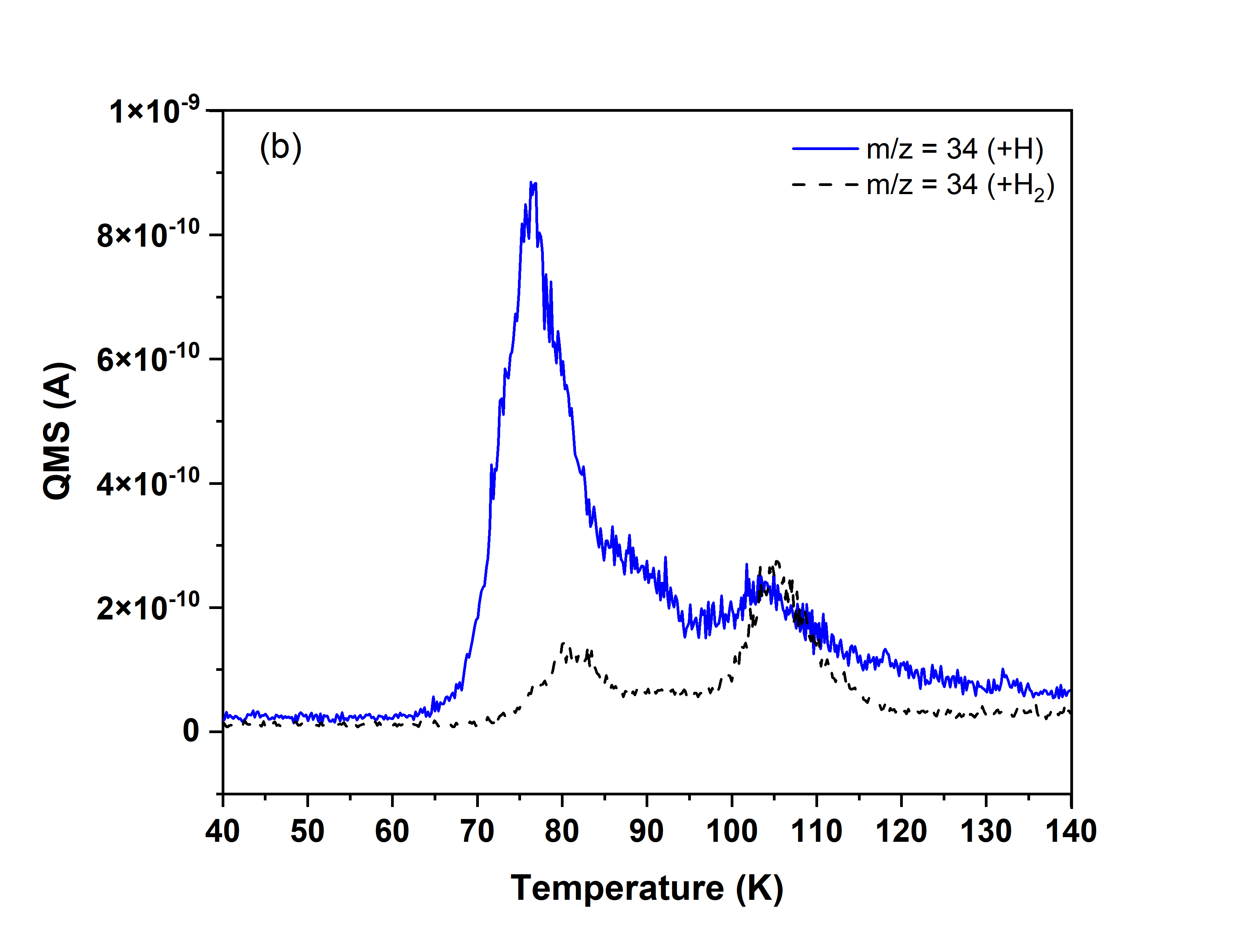}
    \includegraphics[width=1.1\linewidth]{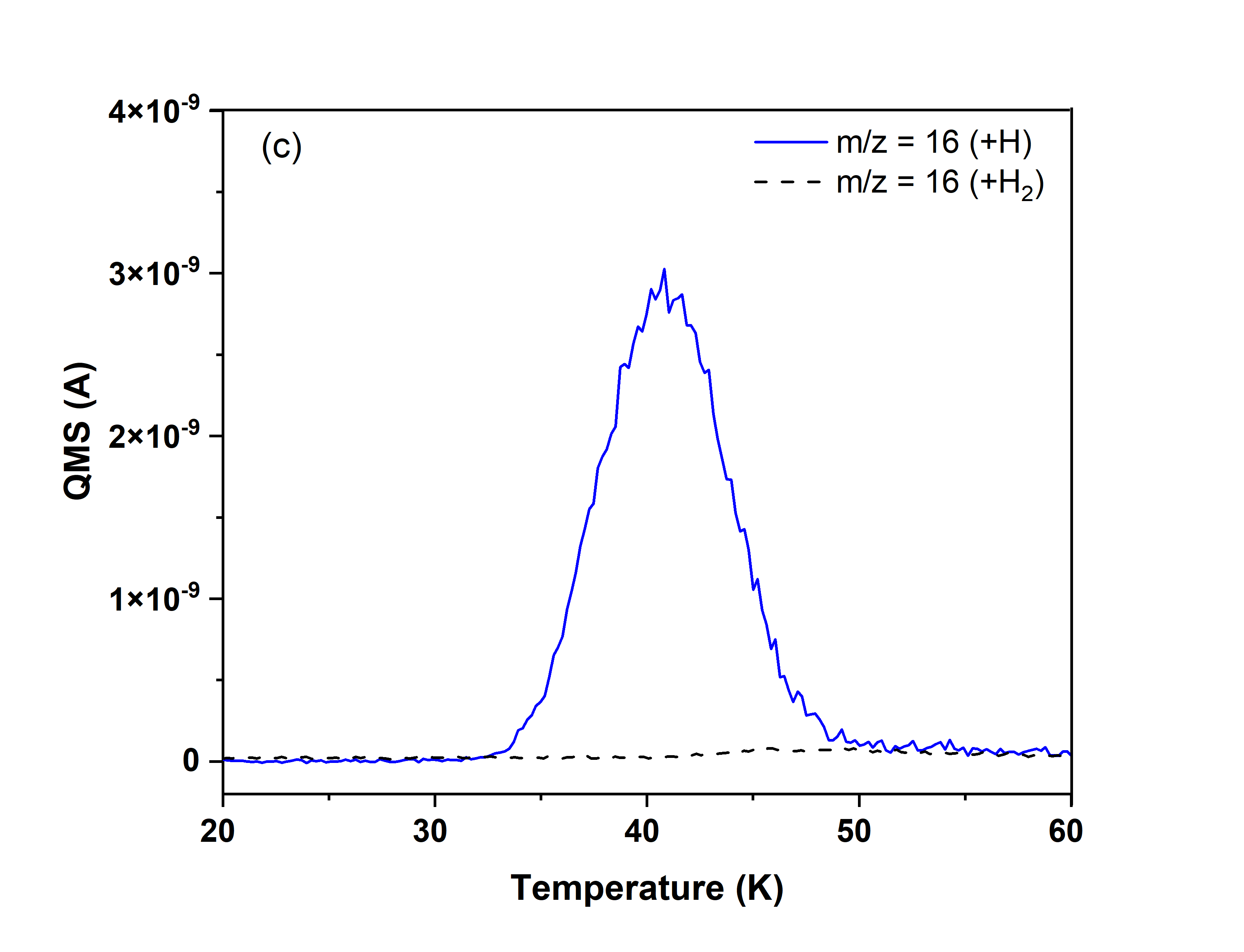}
    \caption{TPD-QMS spectra of the pre-deposited CS$_2$ with H atoms (solid blue lines), compared with H$_2$ molecules (dashed black lines), on c-ASW at 10~K for up to 2 h: (a) at m/z~=~48 (for CH$_3$SH), (b) at m/z~=~34 (for H$_2$S), and (c) at m/z~=~16 (for CH$_4$). }
    \label{QMS-CS2predeposition}
\end{figure}

To verify the presence of these products after the pre-deposition experiments, we monitored the TPD-QMS profiles showing desorption peaks at m/z = 16 (\ce{CH4}), 34 (\ce{H2S}), and 48 (\ce{CH3SH}). The inclusion of \ce{CH4} follows previous results \citep{Nguyen2023}, where it was identified as a fragmentation product of \ce{CH3SH}, as discussed in more detail below. All expected peaks were detected (Figure \ref{QMS-CS2predeposition}). In particular, the desorption peak at m/z = 48, observed between 90 and 115~K, corresponds to CH$_3$SH desorption (Figure \ref{QMS-CS2predeposition}a), in agreement with our earlier findings \citep{Nguyen2023}.

After formation on the icy substrate via H-abstraction reaction of SCH$_2$SH, H$_2$S can subsequently react with H atoms, leading to the loss of H$_2$S from the ice surface via chemical desorption \citep{Oba2018, Oba2019}.
Figure \ref{QMS-CS2predeposition}b shows the desorption peak at m/z~=~34 observed in the temperature in range of 80~-~95~K, which is attributed to H$_2$S remaining on the surface. The second peak, observed at approximately 95~-~130~K, corresponds to the fragment ($^{34}$S) from C$^{32}$S$^{34}$S. In addition, a small desorption peak (dashed black line in the H$_2$ case) at m/z~=~34 in the temperature range of 75~-~85~K may indicate the contamination from the beam line. The production of H$_2$S from the reaction of CS$_2$ with H atoms is confirmed by the desorption peak at m/z~=~34 in the TPD experiments. Figure \ref{QMS-CS2predeposition}c displays the desorption peak observed at m/z~=~16 in the temperature range of 30~-~50~K, corresponding to the typical desorption temperature of CH$_4$, which is likely to proceed through the reaction of H atoms with CH$_3$ radical. These CH$_3$ radicals are yielded through CH$_3$SH with H atoms with low activation barriers or barrierless on c-ASW at 10~K following the reactions pathways by CH$_3$SH~+~H~$\xrightarrow{}$~CH$_3$~+~H$_2$S and CH$_3$~+~H~$\xrightarrow{}$~CH$_4$ \citep{Qasim2020, Nguyen2023}. Additionally, the H atoms landed on dust grains can efficiently consume trapped radicals and unsaturated species through addition and abstraction reactions, leading to smaller, stabilized products rather than larger species \citep{Oberg2009_2}, as supported by our computational results. In contrast, CS$_2$ polymerization typically requires energetic processing, such as UV irradiation or high-energy electrons \citep{CATALDO1995, Heymann2000}. Therefore, CS$_2$ polymers cannot form under our present experimental conditions.

The absence of observed IR features corresponding to HC(S)SH, CH$_3$SH, and H$_2$S from the reaction of pre-deposited CS$_2$ with H atoms is likely due to collisions and subsequent reactions of the products with other species (e.g., H atoms) on the substrate. These interactions may result in either insufficient accumulation of new species or chemical desorption of products into the gas phase, making them undetectable by FTIR. Additionally, the small amount of pre-deposited CS$_2$ ($\sim$1~ML) and the backward reaction that reforms CS$_2$ during the second hydrogenation step may further contribute to the lack of IR-detectable products. In order to shed further light into the products of the title reaction we perform additional experiments using co-deposition, indicated in Section \ref{sec:co-deposition-H}.

\subsubsection{Identification of products through co-deposition of CS$_2$ with H atoms on c-ASW at 10~K} \label{sec:co-deposition-H}

\begin{figure}
    \centering
    \includegraphics[width=\linewidth]{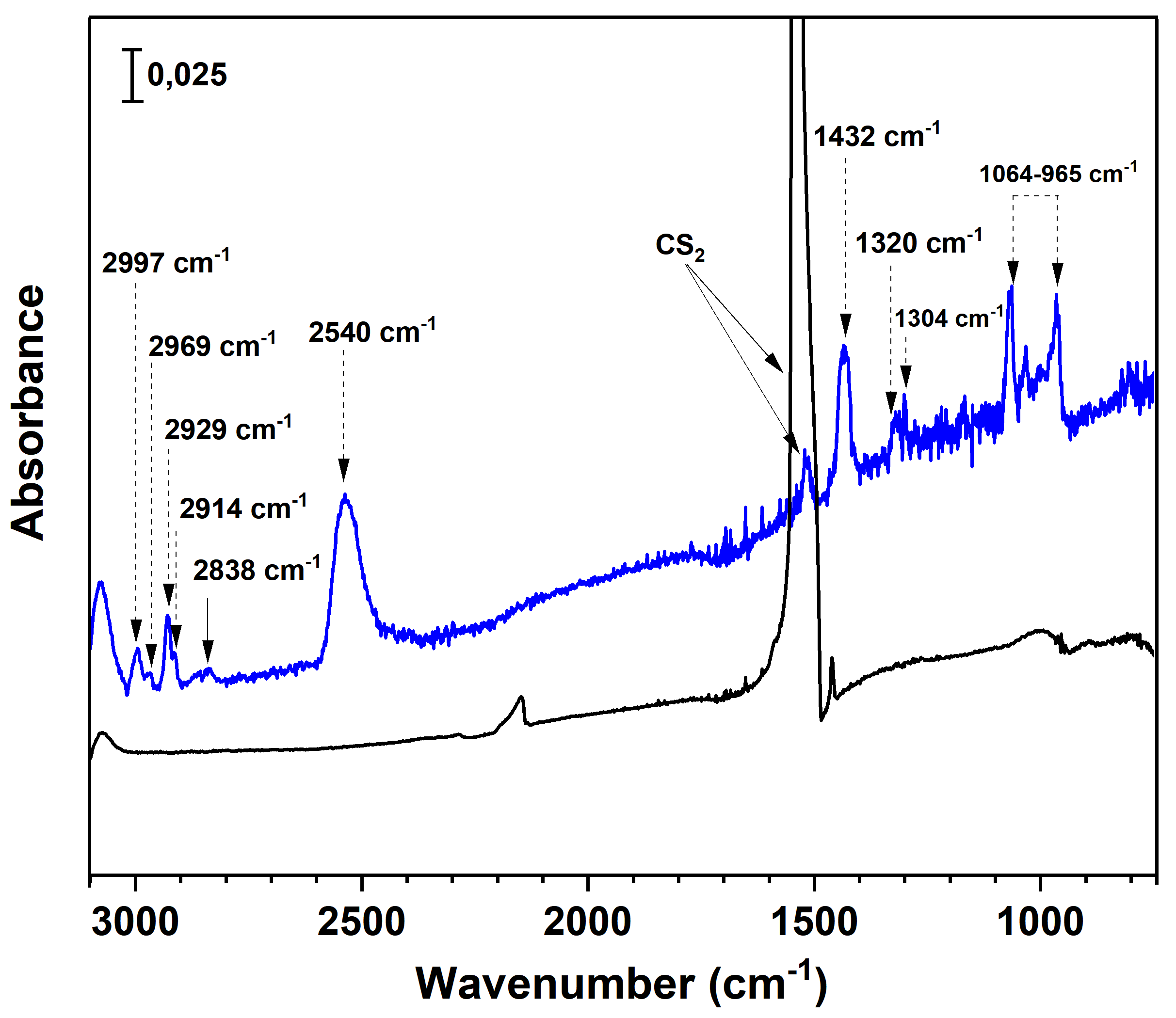}
    \caption{FTIR spectrum of the co-deposited CS$_2$ with H atoms (blue line) after 2 hours with a low deposition rate of 0.08~ML.min$^{-1}$ compared to the co-deposited CS$_2$ with H$_2$ (black line) on c-ASW at 10~K. The dashed arrows present the new features formed on the ice surface, they could be derived from the S-bearing species such as H$_2$S, CH$_3$SH, H$_2$CS, and CH$_4$. The solid arrows presents remaining CS$_2$ after 2 hours of the co-deposition.}
    \label{fig:codepositionCS2+H_LDR}
\end{figure}

In Figure (\ref{fig:codepositionCS2+H_LDR}), we show the FTIR spectrum of solid CS$_2$ (blue line) after co-deposition with H atoms for 2-hours on c-ASW at 10~K, and that with H$_2$ as a blank (back line) for comparison.The deposition rate of \ce{CS2} was 0.08~ML min$^{-1}$. The co-deposition of CS$_2$ with H atoms resulted in the appearance of several new peaks, which were not observed in the blank. 

Notably, several new bands appeared in the regions of approximately 3000–2800 cm$^{-1}$ and 1500–900 cm$^{-1}$ regions (see dashed arrows in Figure~\ref{fig:codepositionCS2+H_LDR}). These were reasonably assigned to methyl mercaptain (CH$_3$SH), based on the following vibrational bands: CH$_3$ antisymmetric stretch (2997 cm$^{-1}$), CH$_3$ symmetric stretch (2929 cm$^{-1}$), CH$_3$ antisymmetric deformation (1432 cm$^{-1}$), CH$_3$ symmetric deformation (1320~cm$^{-1}$), and CH$_3$ rocking modes (1064–965~cm$^{-1}$)  \citep{Hudson2016, Nguyen2023}.

Although \ce{CH3SH} was not included in the possible reaction pathways shown in the previous section, it is likely that methanedithiol (\ce{CH2(SH)2}) may react with H atoms to yield \ce{SCH2SH} radicals (E$_a$ = 11.4 kJ mol$^{-1}$), which would further react with H atoms with a negligible barrier (1.1 kJ mol$^{-1}$), resulting in the formation of H$_2$S and H$_2$CS (reaction \ref{sch2sh_h}). The hydrogenation of H$_2$CS, formed by reaction (\ref{sch2sh_h}) can yield CH$_3$SH through reaction (\ref{rct:4_CH3SH}), as proposed by \cite{Lamberts2018}. 

An absorption peak at 2540~cm$^{-1}$ observed after the co-deposition was attributed to the S–H stretching band in the products, such as H$_2$S and CH$_3$SH \citep{Fathe2006, Oba2019, Nguyen2023}. Furthermore, weak features at 2969 and 2838~cm$^{-1}$ were assigned to the CH$_2$ symmetric stretching band of H$_2$CS, which has been previously observed in the solid phase \citep{Torres1982, Watanabe1991, Suzuki2007}. The observed shifts in H$_2$CS absorption bands relative to the reference values were probably due to substrate-dependent effects during formation. In addition, a small peak observed at 1304~cm$^{-1}$ was assigned to CH$_4$ \citep{Qasim2020}. 

\begin{figure}
    \centering
    \includegraphics[width=\linewidth]{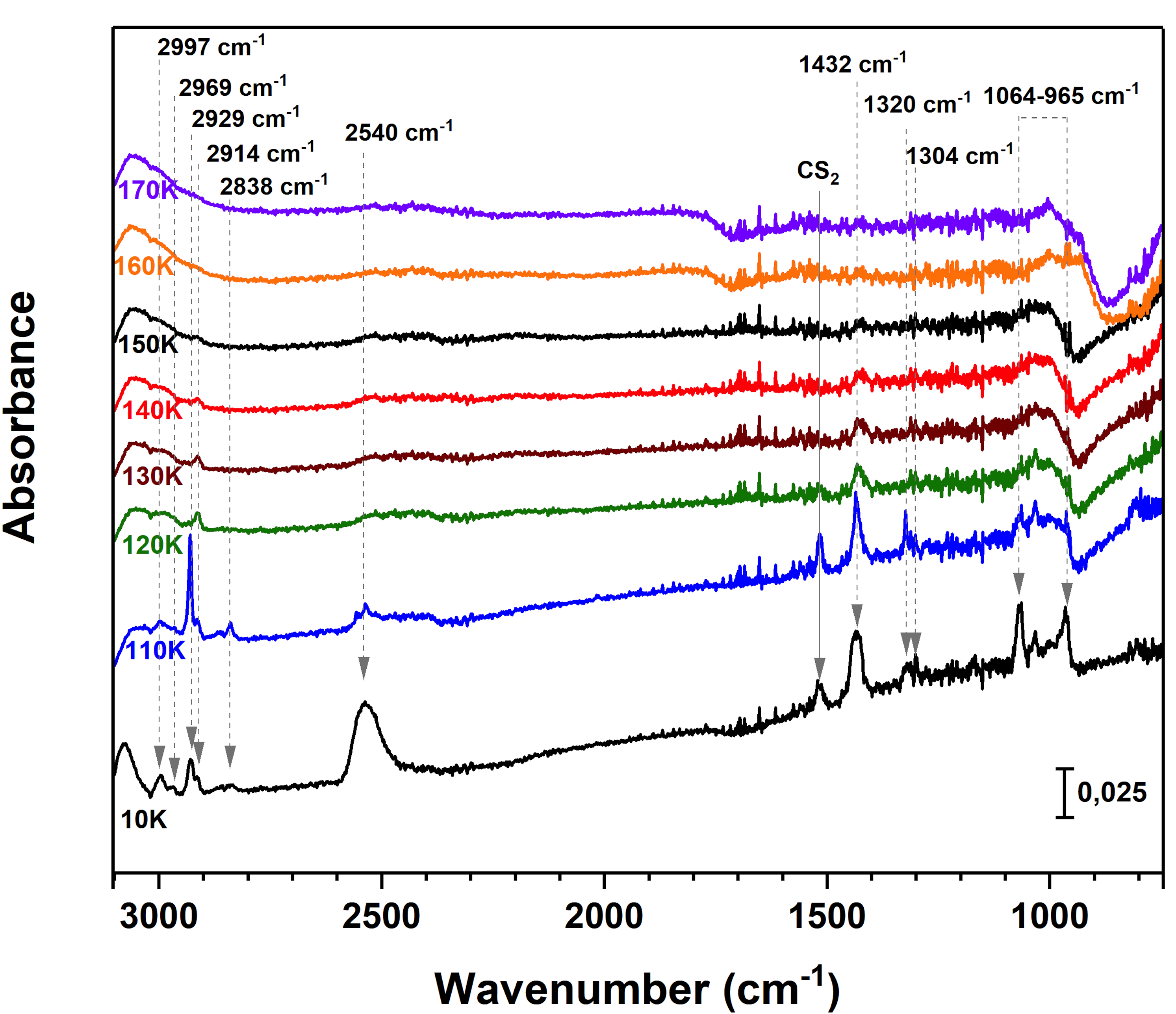}
    \caption{Variation of the different spectra of the sample for co-deposited CS$_2$ with H atoms at 10 K, recorded when the surface temperature was gradually warm up from 10 to 170~K.}
    \label{fig:warmup-FTIR-LPR}
\end{figure}

Figure \ref{fig:warmup-FTIR-LPR} displays the variation in the different spectra of the co-deposited CS$_2$ with H atoms at 10~K, recorded when the surface temperature was gradually warmed up from 10 to 170~K. The S–H stretching band at 2540 cm$^{-1}$ primarily decreases at temperatures below 110~K, suggesting the desorption of H$_2$S as reported by \citet{Oba2019}. The desorption of CH$_3$SH completed in the temperature range of 110~–~125~K \citep{Nguyen2023}, which contributed to the decreases of the peak intensities at 2540, 3000~–~2800, and 1500~–~900~cm$^{-1}$ regions. These desorption temperatures align with the TPD signals observed at m/z~=~34 and 48 (see Figure \ref{QMS-CS2predeposition}). 

\begin{figure}
    \centering
    \includegraphics[width=\linewidth]{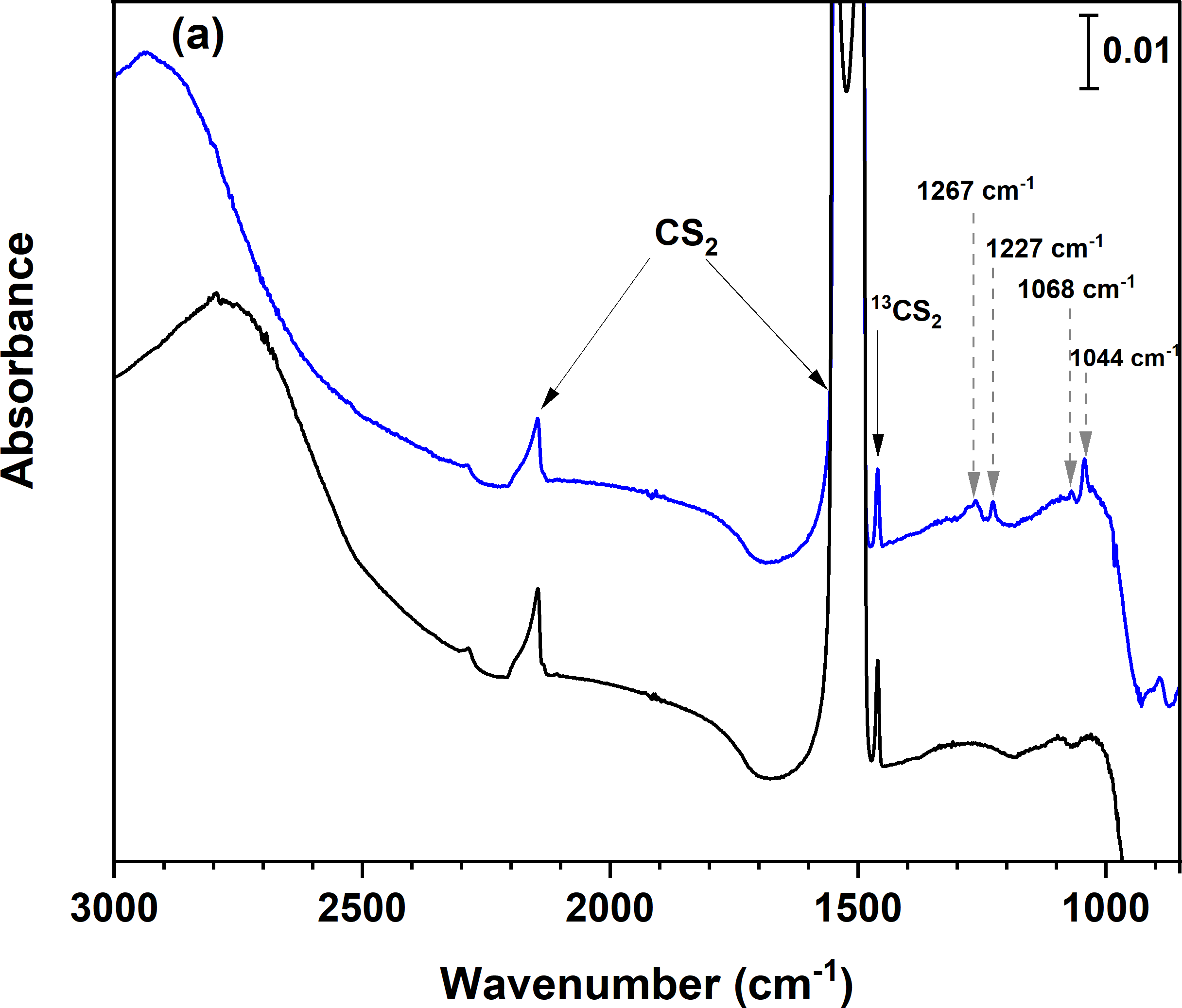}
    \smallskip
    \includegraphics[width=\linewidth]{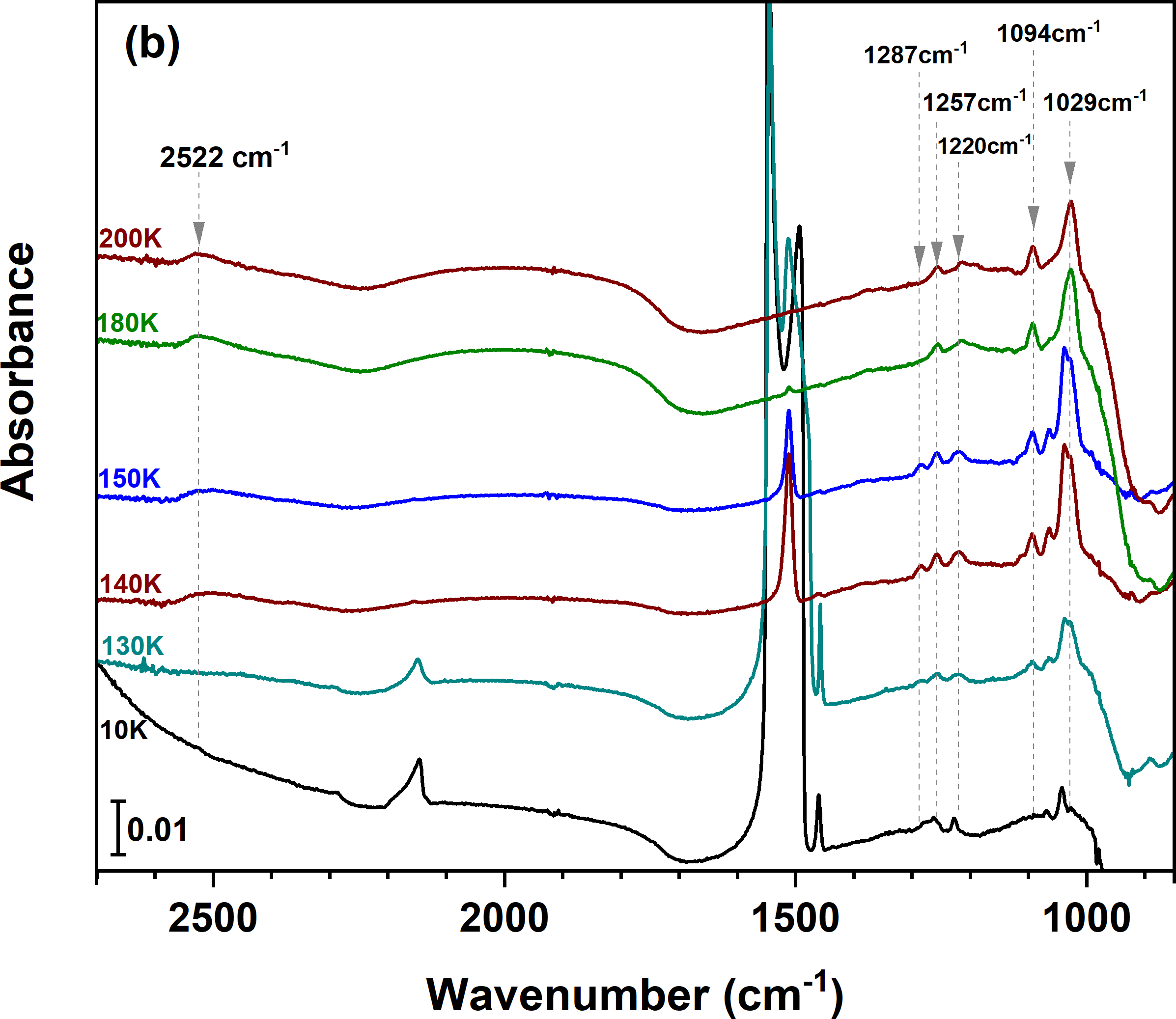}
    \caption{(a) FTIR spectrum of co-deposited CS$_2$ with H atoms (blue curve) after 2 hours under high deposition rate of 1~ML min$^{-1}$ compared with the co-deposition of CS$_2$ with H$_2$ (black curve) as a reference on c-ASW at 10~K. The dash arrows indicate the new features yielded through the reaction of CS$_2$ and H atoms. The solid arrows present remaining CS$_2$ on the substrate after 2 hours co-deposition with H/H$_2$. (b) Variation in the FTIR spectrum of products after the co-deposition of CS$_2$ with H atoms as the surface temperature was warmed up from 10 to 200~K. }
    \label{fig:FTIR-codepositionCS2_HDR}
\end{figure}

Figure \ref{fig:FTIR-codepositionCS2_HDR}a shows the FTIR spectrum of CS$_2$ co-deposited with H atoms at a high deposition rate, of CS$_2$ $\sim$1ML min$^{-1}$, which is $\sim$13 times higher than the previous experiments. By changing the deposition rate of CS$_2$ (that of H was fixed at the constant rate), the absolute and relative yields of products can vary depending on the relative flux of the co-deposited species. For example, when oxygen molecules (O$_2$) and H atoms were co-deposited onto a reaction substrate at 10~K, hydrogen peroxide (H$_2$O$_2$) and water (H$_2$O) were formed as products through the following pathways: 
\begin{equation}
    \ce{O2 -> HO2 -> H2O2 -> H2O + OH}.
\end{equation}
where an arrow indicates an addition of one H atom to the reactant. We have experimentally confirmed that the H$_2$O/H$_2$O$_2$ ratio in the products decreased with increasing the co-deposited O$_2$/H ratio \citep{Oba2009}. This is interpreted as follows: with increasing the O$_2$/H ratio, the number of H atoms which can react with formed H$_2$O$_2$ should decrease, resulting in the formation of less H$_2$O, that is, a low H$_2$O/H$_2$O$_2$ ratio in the product. Hence, when CS$_2$ was co-deposited with H atoms at higher flux, we expect that some reaction intermediates such as HC(S)SH and CH$_2$(SH)$_2$, which can form at an earlier stage of the reaction sequence (Figure \ref{fig:summary}), can remain intact in the reaction products. Indeed, new characteristic bands appeared on the ice substrate (Figure \ref{fig:FTIR-codepositionCS2_HDR}a), which were not observed at lower CS$_2$ deposition rates (see Figure \ref{fig:codepositionCS2+H_LDR}). 

After the co-deposition at the high CS$_2$/H ratio for 2 hours, new absorption peaks appeared in the 1270~-~1220~cm$^{-1}$ and 1100~-~1000~cm$^{-1}$ regions, which are likely to be attributed to sulfur-bearing products. Figure \ref{fig:FTIR-codepositionCS2_HDR}b displays the variation in the FTIR spectra with increasing the substrate temperature. As the surface temperature increases, new absorption features appear, likely due to the accumulation of products in the bulk CS$_2$ after its desorption.
The assignment of the absorption peak at 1287 cm$^{-1}$, which gradually increased with temperature up to 180~K, might be attributed to the C-H rocking vibrational band of HC(S)SH \citep{Lignell2021}. The band at 1044~cm$^{-1}$ increased in intensity upon warming to 180~K (Figure \ref{fig:FTIR-codepositionCS2_HDR}b), supporting its assignment to the C=S stretching mode of HC(S)SH \citep{Bohn1992, Lignell2021}. According to our quantum chemical calculations, HC(S)SH can be formed through reaction (\ref{eq:h-hscs}). 

\begin{figure}
    \centering
    \includegraphics[width=1.02\linewidth]{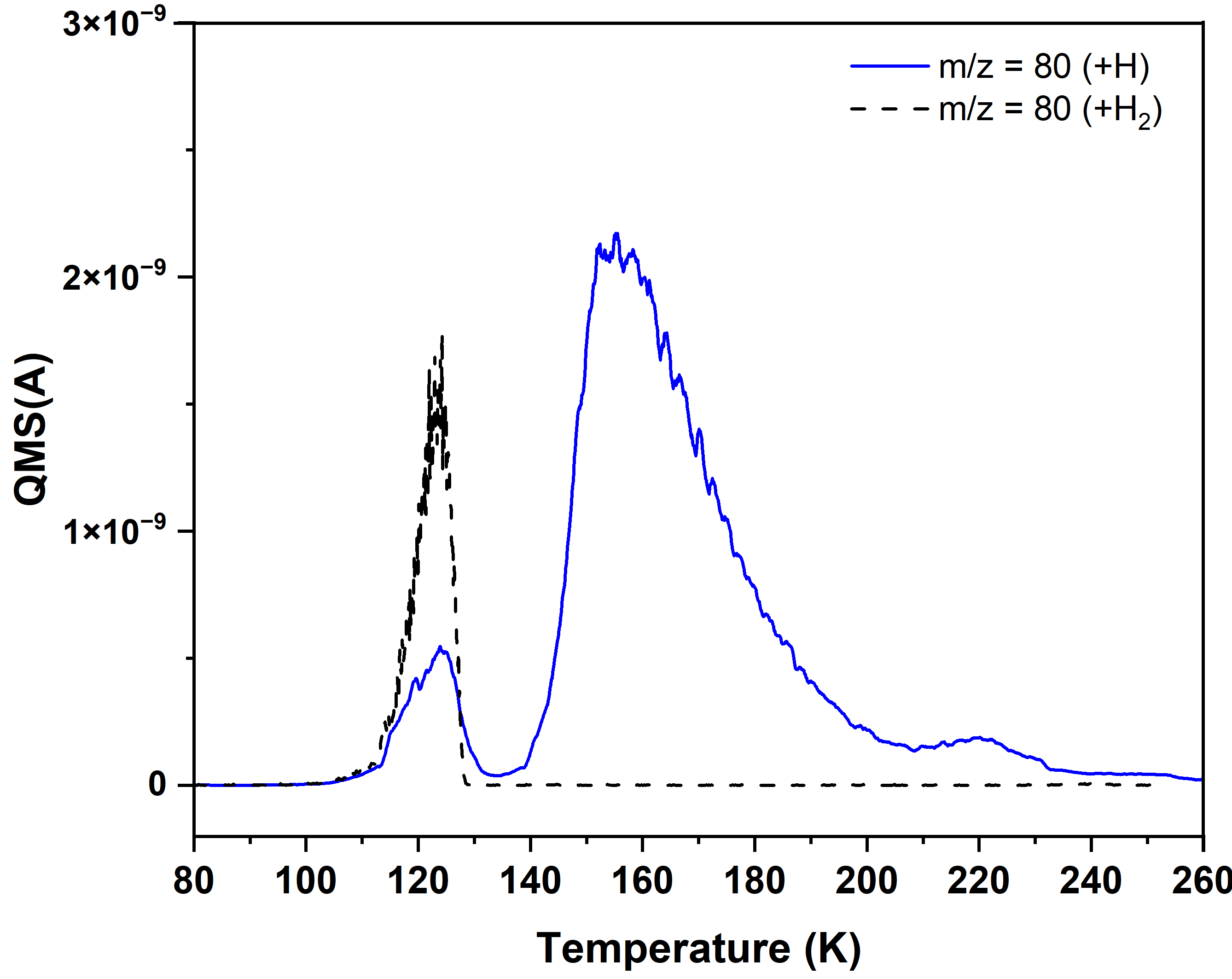}
    \caption{TPD-QMS profile of the desorption peak at m/z = 80 (for CH$_2$(SH)$_2$) after co-deposition of CS$_2$ with H atoms (solid blue line), compared with H$_2$ (dashed black line), at 10~K for up to 2 h.}
    \label{fig:TPD_M80}
\end{figure}

Figure \ref{fig:FTIR-codepositionCS2_HDR}b also shows an absorption peak at 1029 cm$^{-1}$, which gradually appeared above 130 K during surface warm-up. In addition, a larger peak at 2522 cm$^{-1}$, emerging above 140 K, may also represent the S-H strength band of S-bearing products. Since H$_2$S and CH$_3$SH cannot remain on the substrate above 130 K, these absorption features are likely attributable to other S-H-bearing species. Moreover, the desorption temperatures of these bands were consistent with the desorption peak at m/z = 80 (see Figure \ref{fig:TPD_M80}), suggesting that they may be tentatively assigned to CH$_2$(SH)$_2$, which can be formed through reaction (\ref{eq:ch2sh2}).
Unfortunately, due to lack of reference data for the infrared feature of CH$_2$(SH)$_2$, the decisive identification of this molecules was not possible in the present study. Nevertheless, based on the computational and experimental results presented here, we strongly expect that CH$_2$(SH)$_2$, as well as HC(S)SH, is one of the possible candidates for the observed peaks even at high temperatures.



\begin{table*}
\begin{center}
    \caption{Infrared Absorption band after co-deposition of CS$_2$ with H atoms on c-ASW}
    \label{tab:IRspectra}
    \begin{tabular}{c c c}
    \toprule 
    Band position (cm$^{-1}$) & Assignment &  References\\
    \midrule
     1064 - 965 &\ce{CH3SH} (\ce{CH3}- rocking) & \citet{Hudson2016}\\
     \midrule
     1044 - 1029 & HC(S)SH (C-S str) & \citet{Bohn1992, Lignell2021} \\
     \midrule
      1287 & HC(S)SH (C-H) & \citet{Lignell2021}\\
     \midrule
     1304 &\ce{CH4} (\ce{C-H} str) & \citet{Qasim2020, Nguyen2023}\\
     \midrule
     1320 &\ce{CH3SH} (-\ce{CH3} symm def) & \citet{Hudson2016}\\
     \midrule
     1442 & \ce{CH3SH} (-\ce{CH3} bending) & \citet{Hudson2016} \\
     \midrule
     1540 &  \ce{CS2} (C=S str) & \citet{Garozzo2008, Edridge2010}\\
     \midrule
     2550 & \ce{H2S}; \ce{CH3SH} (H-S str) & \citet{Fathe2006, Hudson2016, Oba2019, Nguyen2023} \\
     \midrule
     2838 & \ce{H2CS} & \citet{Torres1982, Watanabe1991, Suzuki2007} \\
     \midrule
     2914 & unknown & \\
     \midrule
     2929 & \ce{CH3SH} (-\ce{CH3} symm str) & \citet{Hudson2016, Nguyen2023} \\
     \midrule
     2969 & \ce{H2CS} (\ce{CH2}- symm str) & \citet{Torres1982, Watanabe1991, Suzuki2007} \\
     \midrule
     2997 & \ce{CH3SH} (-\ce{CH3} antisymm str) & \citet{Hudson2016, Nguyen2023} \\
     \bottomrule
    \end{tabular}
\end{center}
\end{table*}

\subsubsection{Comparison of the CS$_2$ depletion after exposure H and D atoms on c-ASW at 10 K} \label{sec:HvsD}
\begin{figure}
   \centering
   \includegraphics[width=\linewidth]{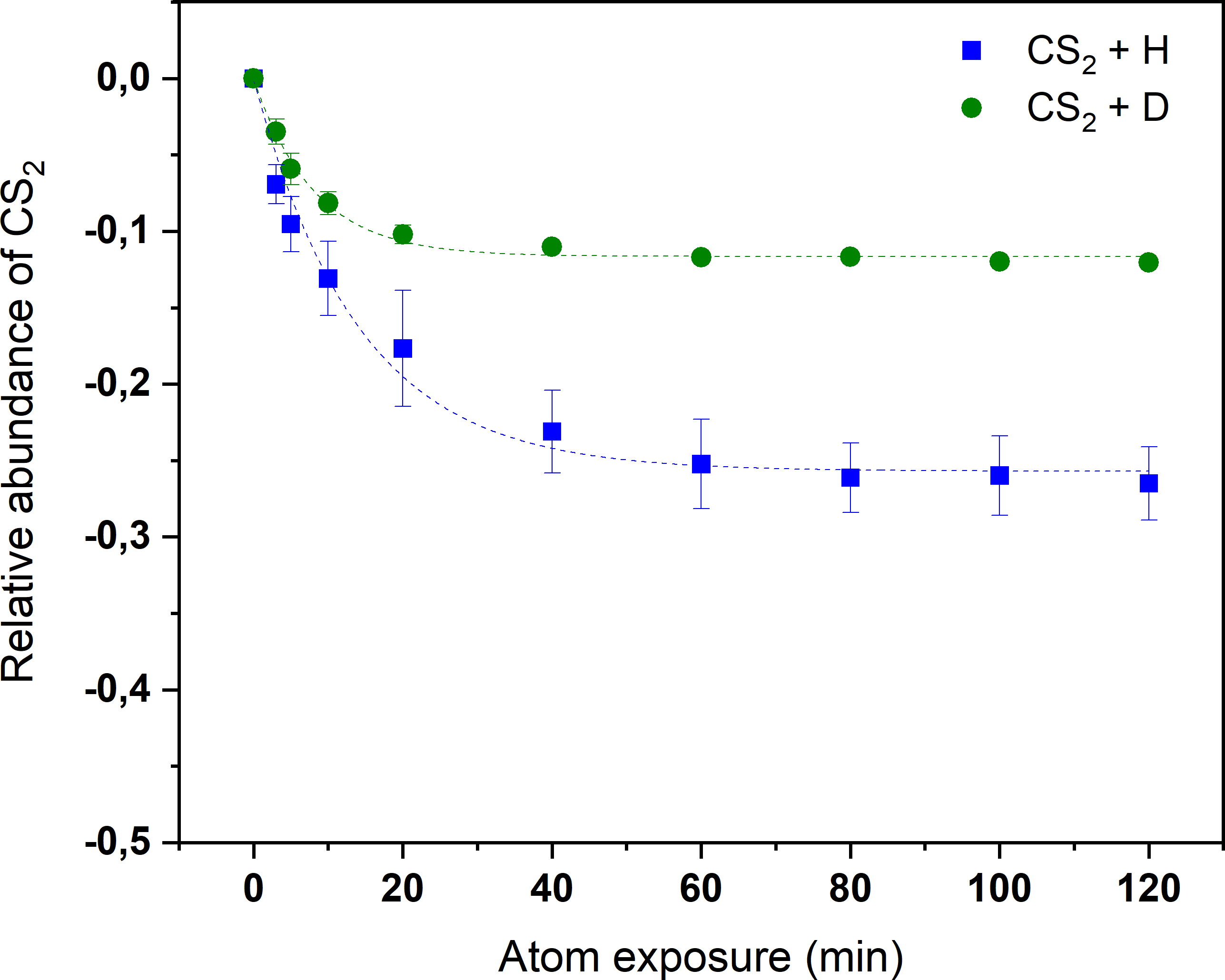}
    \caption{Relative abundance of solid CS$_2$ on c-ASW with exposure to H (blue squares) and D (green circles) atoms at 10~K. The dash lines are the fits to Equation (\ref{eq:2}).}
    \label{fig:CS2consumption_HvsD}
\end{figure}

In Sections \ref{sec:comp} and \ref{sec:exp}, we anticipate that, in order to activate \ce{CS2}, that is, to first hydrogenate the molecule and move \textit{forward} in the reaction network, the system should overcome an activation energy of 13.1 \kjmol corresponding to reaction (\ref{eq:h-cs2}). To confirm such an expectation, we determined the kinetic isotopic effect (KIE) of the title reaction, that, as we see in Figure \ref{fig:summary}, is mostly dominated by the first hydrogenation. Figure \ref{fig:CS2consumption_HvsD} presents the change in the relative abundance of solid CS$_2$ after exposure to H and D atoms on c-ASW at 10~K. After up to 2 hours of exposure, the fractional loss of CS$_2$ was approximately 26$\%$ and 12$\%$ for exposure to H and D atoms, respectively. 
The loss of CS$_2$ is caused by the reaction with H (or D) atoms only; therefore, assuming that the reaction of CS$_2$ with H (or D) atoms proceeds through a single process where we merge all possible reactions, and the relative abundance of CS$_2$ can be represented by the following rate equation: 
\begin{equation}
       \mathrm{d[CS_2]}/dt = -k_\mathrm{X}\mathrm{[X][CS_2]}
    \label{eq:1}
\end{equation}
where [CS$_2$] and [X] are the surface number densities of CS$_2$ and X atoms (X = H or D) on the ice surface, respectively, and \textit{k}$_\textrm{X}$ is the rate coefficient for the X addition reactions. The reaction kinetics can be derived using the following integrated equations (\ref{eq:2}): 
\begin{equation}
   \rm{\Delta[CS_2]}_t/\rm{[CS_2]_0} = A \times (1 - exp(-k^{'}_{\rm{X}}t)),
    \label{eq:2}
\end{equation}
where $\Delta$[CS$_{2}$]$_{t}$ is the change in the abundance of the solid CS$_{2}$ at time t, and [CS$_{2}$]$_{0}$ is the initial CS$_{2}$ abundance. {\textit{A}} is the saturation value for the decrease in CS$_{2}$, and {\textit{k}}$^{'}_{\textrm{X}}$ (= {\textit{k}}$_{\textrm{X}}$[X]) is the effective rate constant for the reaction of CS$_{2}$ with X atoms. By fitting the data plot in Figure \ref{fig:CS2consumption_HvsD} using equation (\ref{eq:2}), we obtained the {\textit{k}}$^{'}_{\textrm{H}}$~=~(1.2~$\pm$~0.1)~$\times$~10$^{-3}$~s$^{-1}$ and {\textit{k}}$^{'}_{\textrm{D}}$~=~(2.0~$\pm$~0.1)~$\times$~10$^{-3}$~s$^{-1}$. According to the previous study of \cite{Kuwahata2015}, when the flux of atoms was on the order to 10$^{14}$ atoms cm$^{-2}$ s$^{-1}$, the number densities of H and D atoms ([D]/[H]) on porous amorphous solid water and crystalline water ices was $\sim$4 and $\sim$10, respectively. By simply assuming that the [D]/[H] on c-ASW is the value between 4 and 10, for example 7, we estimate the ratio of {\textit{k}}$_{\textrm{D}}$/{\textit{k}}$_{\textrm{H}}$ from the {\textit{k}}$^{'}_{\textrm{D}}$/{\textit{k}}$^{'}_{\textrm{H}}$, following the relationship:

\begin{equation}
    \frac{k_{\rm{D}}}{k_{\rm{H}}} = \frac{k^{'}_{\rm{D}}}{k^{'}_{\rm{H}}} \times \frac{[\rm{H}]}{[\rm{D}]}
   = 1.7 \times \frac{1}{7} 
    = 0.24
    \label{eq:3}
\end{equation}
Based on the assumption, the ratio of \textit{k}$_D$/\textit{k}$_H$ obtains a value of 0.24, indicating that the D addition reaction of solid CS$_2$ is slower than the H addition reaction. This fraction, or more precisely its inverse ($\sim 4.2$), represents the KIE, which, in the case of these reactions, is attributed to the influence of quantum tunneling. Although this KIE is relatively small, it is still noticeable, confirming that the reaction efficiency can be attributed to this effect. However, we note that it is derived from the overall decay of the \ce{[CS2]_H}/\ce{[CS2]_D} band, rather than from individual reaction pathways. 


\section{Astrophysical Implications} \label{sec:astro}

The target molecule of this study, CS$_2$ has so far played a more significant role in cometary ices than in interstellar ices, as it has not been detected in interstellar environments. In fact, the astronomical detection of CS$_2$ in the gas phase is completed by the fact that it is a non-polar molecules, while detection in ices via IR vibrational bands has so far proven unsuccessful. Despite this hardships, models predict CS$_2$ can reach a total (gas + ice) fractional abundance of $\sim$10$^{-10}$ \citep{Laas2019}. In fact, CS$_2$ is likely to be destroyed through reactions with atoms such as oxygen and/or hydrogen on dust grains. In particular, reactions between \ce{CS2} with O at thermal conditions can yield OCS on interstellar dust grains \citep{Ward2012}. In the case of reactions with atomic hydrogen, which are the purpose of this study, CS$_2$ can react through reactions (\ref{eq:h-cs2}) - (\ref{sch2sh_h}), CS$_2$ will be consumed on dust grains, leading to the formation of other S-bearing species, including CH$_3$SH, H$_2$S, HC(S)SH, and CH$_2$(SH)$_2$ (see Table \ref{tab:summary}). If chemical routes of CS$_2$ proceed efficiently on dust grains, the production of S-bearing species would be controlled by the abundance of hydrogen atoms available to react with CS$_2$ molecules. Destruction reactions produce CS/\ce{H2CS} and H$_2$S, resulting in the cleavage of the original molecular skeleton. The CS/\ce{H2CS} molecules are then directly related with CS$_2$ and would be hydrogenated to yield CH$_3$SH \citep{Lamberts2018}. Meanwhile, a fraction of H$_2$S will be released into the gas phase by chemical desorption \citep{Oba2018}. Previous studies \citep{Nguyen2021, Molpeceres2021b, Nguyen2023, Nguyen2024} demonstrated that the hydrogenation of key S-bearing species, including OCS, CH$_3$SH, SO$_2$, and CS$_2$, lead to the formation of H$_2$S. The efficient conversion of sulfur species, including organosulfur compounds, to \ce{H2S} might be one of the, surely multicausal, explanations for the non-detection of S-bearing species in the solid state.  

A possible way to provide some insights on CS$_2$ in the gas phase could be obtained from the observation of its protonated species, HSCS$^{+}$. Unfortunately, this species has a very low dipole moment along the a-axis of 0.19D. Moreover, only transitions of a-type have been observed in the laboratory \citep{McCarthy2009}. Although, a larger dipole moment along the b-axis is expected, the frequencies of the b-type transitions remain unknown. From the a-type transitions, and using the ultrasensitive line survey \textsc{QUIJOTE} \citep{Cernicharo2021}, we obtain a 3 sigma upper limit to the abundance of HSCS$^{+}$ in the gas phase of 2 $\times$ 10$^{12}$~cm$^{-2}$ for each line in
the survey. Averaging of the lines J=5-4 to J=8-7, the 3 sigma upper limit to the column density of CS$_2$ is 1 $\times$ 10$^{12}$~cm$^{-2}$. Protonated species have abundances between 50 and 100 times lower than the corresponding neutrals. Hence, the upper limit to CS$_2$ in the gas-phase is not very constraining in TMC-1, of the order of 50-100 $\times$ 10$^{12}$~cm$^{-2}$, which is a consequence of the low dipole moment of the protonated molecule.

The hydrogenation of \ce{CS2} on interstellar dust grains leads to the formation of two molecules that can be considered more complex than the parent species: \ce{HC(S)SH} and \ce{CH2(SH)2}. It is therefore reasonable to explore whether these molecules may be present in the interstellar medium (ISM). Although current spectroscopic data are insufficient to enable the detection of \ce{CH2(SH)2}, the presence of \ce{HC(S)SH} has recently been reported by \citet{Manna2024} in the hot core NGC 1333 IRAS 4A2. We have searched for \ce{HC(S)SH} in TMC-1 using data from the \textsc{Quijote} line survey \citep{cernicharo_quijote_2022}. These observations yielded no detection, as detailed in Appendix~\ref{observations}. The non-detection of \ce{HC(S)SH} can be attributed to several factors, but is nevertheless consistent with our findings, particularly the small energy barrier calculated for Reaction~\ref{eq:h-hcssh}. This low barrier facilitates the rapid hydrogenation of \ce{HC(S)SH}, leading to its efficient destruction. This interpretation is supported by our experimental results, where \ce{HC(S)SH} is only observed in co-deposition experiments and only in small quantities. At higher temperatures, such as those characteristic of hot cores, the hydrogenation of \ce{HC(S)SH} may proceed even in the gas phase, potentially yielding bimolecular products. This hypothesis warrants further investigation through laboratory experiments or accurate rate constant calculations. Similar considerations apply to \ce{CH2(SH)2}. Although the reaction barrier for its hydrogenation is larger and its destruction in the gas phase may not be as efficient, hydrogenation on icy grain surfaces is expected to proceed rapidly. This process likely forms \ce{SCH2SH} (see Reaction~\ref{eq:ch2sh2_h}) and ultimately yields \ce{H2S} and \ce{H2CS} after a final hydrogenation step. These findings convey a twofold message. First, we demonstrate that hydrogenation of \ce{CS2} efficiently destroys the S=C=S functional group. Second, the hydrogenation of not only \ce{CS2} but also its reaction intermediates raises serious doubts about the long-term stability of the hydrogenation products \ce{HC(S)SH} and \ce{CH2(SH)2} in the ISM. In particular, additional observational and modelling  efforts are required to confirm the recent detection of \ce{HC(S)SH} toward NGC 1333 IRAS 4A2 \citep{Manna2024}.

\section{Conclusion}

The chemical pathways of CS$_2$ with H atoms in interstellar ices at low temperatures have been investigated. The hydrogenation of CS$_2$ leads to multiple reaction channels, including the formation of H$_2$S and CH$_3$SH, as well as HC(S)SH and CH$_2$(SH)$_2$. According to quantum chemical calculations, these species can form on dust grain surfaces via low-barrier or barrierless reactions, with their formation efficiencies dependent on the local abundance of CS$_2$. The non-detection of CS$_2$ in interstellar ices can therefore be attributed to its efficient destruction. In regards the formation of HC(S)SH and CH$_2$(SH)$_2$ compounds from \ce{CS2} under ISM conditions, the extrapolation of our experimental and theoretical results suggest that these molecules should not withstand a wide array of standard chemical conditions in the ISM, including cold and hot cores.

\section*{Acknowledgements}
This work was partly supported by JSPS KAKENHI grant Nos. JP25H00677, JP23H03980, JP21H04501 (to Y.O.), JP22H00159 (to N.W.), and JP21F21319 (to T.N., N.W.).
GM acknowledges the support of the grant RYC2022-035442-I
funded by MICIU/AEI/10.13039/501100011033 and ESF+. GM
also received support from project 20245AT016 (Proyectos Intramurales CSIC). We acknowledge the computational resources provided the DRAGO computer cluster managed by SGAI-CSIC, and the Galician Super-
computing Center (CESGA). The supercomputer FinisTerrae III and
its permanent data storage system have been funded by the Spanish
Ministry of Science and Innovation, the Galician Government, and
the European Regional Development Fund (ERDF). MA and JC thank Spanish Ministerio de Ciencia, Innovaci\'on, y Universidades for funding through the project PID2023-147545NB-I00. GE acknowledges support from Spanish grant PID2022-137980NB-I00, funded by MCIN/AEI/10.13039/501100011033/FEDER UE.  We also thank ERC for funding through grant ERC-2013-Syg-610256-NANOCOSMOS. The observations of TMC-1 molecular cloud are carried out with the Yebes 40m telescope (projects 19A003, 20A014, 20D023, 21A011, 21D005, and 23A024). The 40m radio telescope at Yebes Observatory is operated by the Spanish Geographic Institute (IGN; Ministerio de Transportes, Movilidad y Agenda Urbana).


\section*{Data availability}

The underlying data supporting the results of this article will be shared on request.



\bibliographystyle{mnras}
\bibliography{CS2ref} 




\appendix




\section{Search for HCSSH in TMC-1} \label{observations}

We have searched for the trans isomer of dithioformic acid, which is the most stable form, in the cold starless core TMC-1, where a large variety of sulfur-containing molecules have been found \citep{Cernicharo2021a,Cernicharo2024,Agundez2025,Cabezas2025,esplugues_first_2025}. The rotational spectrum of trans HCSSH has been measured in the laboratory \citep{Prudenzano2018} and the dipole moment components along the $a$ and $b$ axes are 1.53 D, as measured by \cite{Bak1978}, and 0.1924 D, as calculated by \cite{Prudenzano2018}. The most favorable lines of trans HCSSH in the Q band (31.0-50.3 GHz) are $a$-type transitions with $K_a$\,=\,0,1. We have searched for them without success in our QUIJOTE data of TMC-1 observed with the Yebes\,40m telescope \citep{Cernicharo2021b}. Assuming a rotational temperature of 9 K, equal to the gas kinetic temperature in TMC-1, an emission size with a diameter of 80$''$, and a line width of 0.60 km s$^{-1}$ \citep{Agundez2023}, we derive a 3$\sigma$ upper limit to the column density of trans HCSSH of 3.5\,$\times$\,10$^{10}$ cm$^{-2}$. Therefore, we could not detect thioformic acid in TMC-1. The HCSSH/HCOOH abundance ratio in TMC-1 is $<$2.5\,$\times$\,10$^{-2}$, similar to the CH$_3$SH/CH$_3$OH and CH$_3$CHS/CH$_3$CHO ratios found in this same source \citep{Agundez2025}.

\begin{figure*}
\centering
\includegraphics[width=0.9\textwidth]{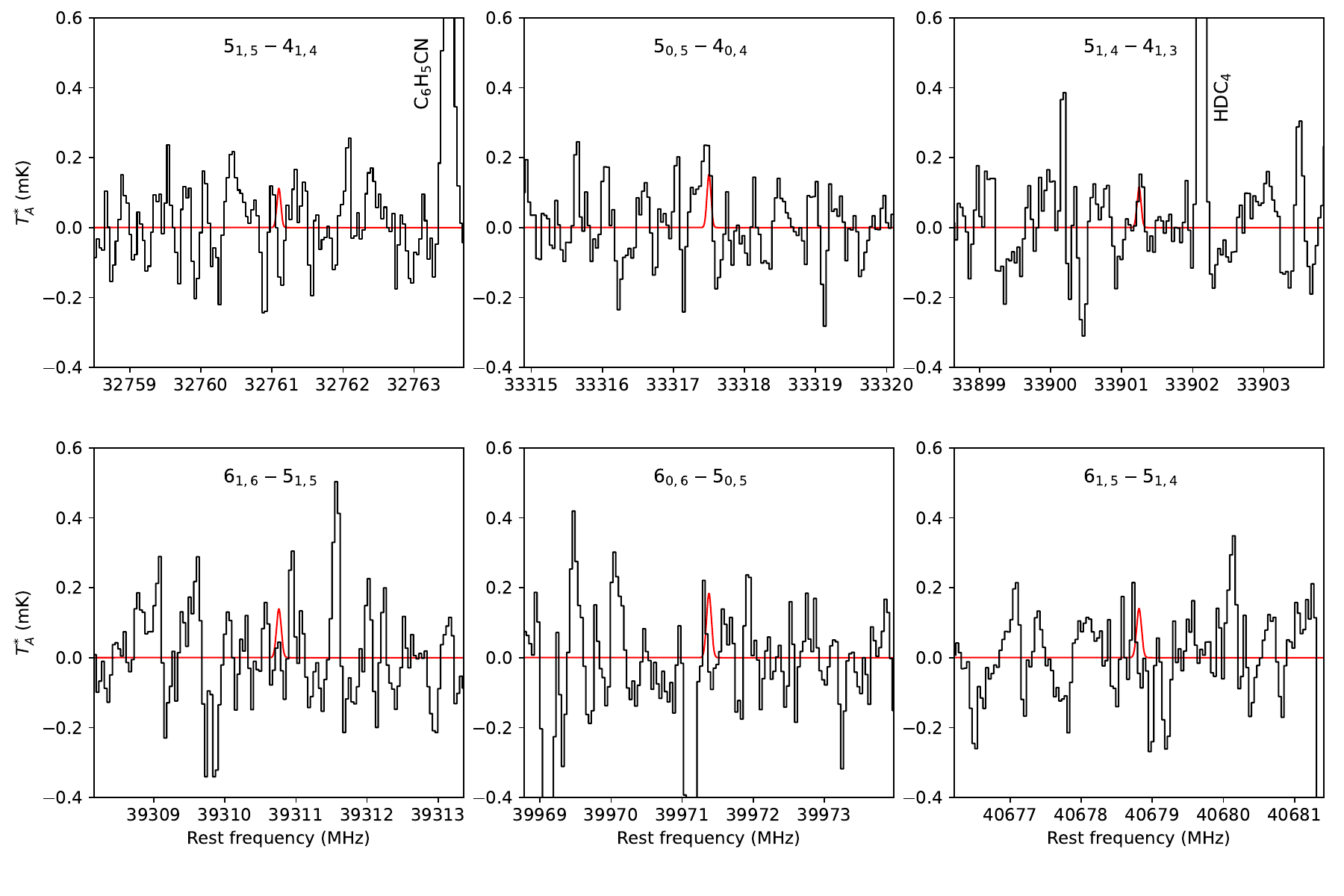}
\caption{Spectra taken with the Yebes 40m telescope toward TMC-1 at the frequencies of the more favorable lines of trans HC(S)SH. We also show in red the calculated spectrum under thermodynamic equilibrium, adopting as column density the 3$\sigma$ upper limit given in the text.}
\label{fig:summary_A}
\end{figure*}


\bsp	
\label{lastpage}
\end{document}